\pdfoutput=1

\documentclass[
 aps,
 prb,
 twocolumn,
 superscriptaddress,
 floatfix,
 longbibliography,
 nofootinbib
]{revtex4-2}

\usepackage{graphicx}
\usepackage{amsmath}
\usepackage{amssymb}
\usepackage{bm}
\usepackage{orcidlink}
\usepackage{chngcntr}
\usepackage{hyperref}
\hypersetup{colorlinks=true,linkcolor=blue,citecolor=blue,urlcolor=blue}

\newcommand{\bk}{\mathbf{k}}
\newcommand{\bG}{\mathbf{G}}
\newcommand{\bQ}{\mathbf{Q}}
\newcommand{\bq}{\mathbf{q}}

\newcommand{\bkbar}{\bar{\bk}}

\begin{document}

\title{Emergence of Topological Electron Crystals in Bilayer
Graphene--Mott Insulator Heterostructures}

\author{Wangqian Miao\,\orcidlink{0000-0002-8182-9528}}
\email{wangqian.miao@connect.ust.hk}
\affiliation{New Cornerstone Science Laboratory, Department of Physics, The Hong Kong University of Science and Technology, Clear Water Bay, Hong Kong, China}
\affiliation{Department of Physics, The Pennsylvania State University, University Park, Pennsylvania 16802, USA}

\author{Tianyu Qiao}
\affiliation{New Cornerstone Science Laboratory, Department of Physics, The Hong Kong University of Science and Technology, Clear Water Bay, Hong Kong, China}

\author{Xue-Yang Song}
\affiliation{New Cornerstone Science Laboratory, Department of Physics, The Hong Kong University of Science and Technology, Clear Water Bay, Hong Kong, China}

\author{Yinghai Xu}
\affiliation{National Laboratory of Solid-State Microstructures, School of Physics, Nanjing University, Nanjing 210093, China}

\author{Yiwei Chen}
\affiliation{National Laboratory of Solid-State Microstructures, School of Physics, Nanjing University, Nanjing 210093, China}

\author{Lei Wang\,\orcidlink{0000-0002-1919-9107}}
\affiliation{National Laboratory of Solid-State Microstructures, School of Physics, Nanjing University, Nanjing 210093, China}
\affiliation{Jiangsu Physical Science Research Center, Nanjing 210093, China}

\author{Xi Dai\,\orcidlink{0000-0002-2396-0966}}
\email{daix@ust.hk}
\affiliation{New Cornerstone Science Laboratory, Department of Physics, The Hong Kong University of Science and Technology, Clear Water Bay, Hong Kong, China}

\date{\today}

\begin{abstract}
The interplay between strong electron correlation and band topology offers a playground for discovering exotic quantum phases. Here, we predict the emergence of topological electron crystals in a charge-transfer bilayer graphene--Mott insulator (BLG--MI) heterostructure. In this system, interlayer charge transfer induces a charge-neutral electron-hole bilayer with \textit{strong mass asymmetry}. While the extreme dilute limit favors a classical triangular dipolar Wigner crystal, we show that increasing the carrier density triggers a critical competition between the Coulomb interaction and the underlying topological band structure of bilayer graphene. This interplay destabilizes the triangular dipolar Wigner crystal and instead stabilizes intrinsic quantum electron crystals with spontaneously formed honeycomb and kagome geometries. Crucially, these new phases can host distinct topological responses including the quantum anomalous and quantum spin Hall effects, which inherit the nonlocal quantum geometry of the bilayer graphene wave functions. Our results establish this artificial heterostructure as a highly tunable platform for mimicking two-dimensional topological solids in a single device.
\end{abstract}

\maketitle

\section{Introduction}

Crystallization is a fundamental phenomenon in condensed matter physics. Beyond atomic systems, various emergent objects and collective degrees of freedom can also form crystalline structures, such as vortices in superconductors, skyrmions in magnetic films, and dimples on the helium surface. Notably, a dilute electron gas can spontaneously crystallize into a Wigner crystal when long-range Coulomb repulsion overwhelms the kinetic energy \cite{wigner_interaction_1934,goldman1990evidence,li2021imaging}. In two dimensions, the energetically favored order is typically triangular, reflecting the classical tendency toward compact packing. Recent progress has shown this conventional picture can be enriched when the crystallizing carriers live in bands with nontrivial Berry curvature and intrinsically nonlocal wave functions. In this setting, charge crystallization can intertwine with band topology, giving rise to topological electron crystals: symmetry breaking charge-ordered states that also exhibit nontrivial topological responses \cite{hall_crystal_prb_1989,zeng_sublattice_2024,tan_parent_2024, soejima_jellium_2025}. Such phases have been recently discussed in rhombohedral graphene multilayers \cite{dong_anomalous_2024,soejima_anomalous_2024,dong_theory_2024, dong_stability_2024, zhou_fractional_2024, zhou_new_2025, desrochers_elastic_2025, guo_correlation_2025, kwan_rmg_prb_2025, yu_rmg_ed_prb_2025,
miao25electron, bab_trashcan_2025}, where quantum anomalous Hall states and their fractional counterparts \cite{lu_fractional_2024,lu_extended_2025, xie_tunable_2025} have been observed.

This raises a sharper question: can an electron--hole bilayer spontaneously form a charge-neutral topological crystal without relying on a moir\'e superlattice or an externally imposed periodic potential \cite{yang_unconventional_2023, sun_signature_2024, zhang_engineering_2024, ghorashi_topological_2023, zeng_gate-tunable_2024, daniel_topo_flat_bands_2024, shi_fractional_2025, zhan_designing_2025, pantaleon_designing_2025,tilak2024proximity, crepel_efficient_2025}? In such a purely self-generated state, the geometry and topology of the order are controlled by interlayer attraction, intralayer Coulomb repulsion, and the quantum geometry of the underlying electronic bands.

\begin{figure}[t]
\centering
\includegraphics[trim=17 16 16 8, clip, width=0.95\columnwidth]{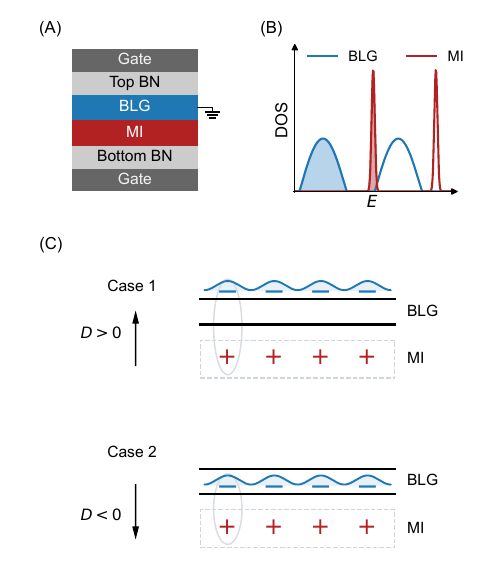}
\caption{\label{fig:setup}
Bilayer graphene--Mott insulator heterostructure.
(A) Schematic of the BN-encapsulated bilayer graphene (BLG)--Mott insulator (MI) heterostructure.
(B) Band alignment in the charge-transfer regime studied in the main text, in which electrons are transferred from the MI to BLG, leaving holes in the MI layer.
(C) The displacement field $D$ opens a gap in BLG and layer-polarizes its low-energy states. Positive (negative) $D$ localizes the electron-doped states predominantly on the upper (lower) graphene layer, so that each transferred electron and the corresponding MI hole form an interlayer dipole (exciton).}
\end{figure}

To address this, we consider a bilayer graphene--Mott insulator (BLG--MI) heterostructure in which charge transfer or modulation doping \cite{arora2025engineering} generates a strongly mass asymmetric electron--hole bilayer. As illustrated in Fig.~\ref{fig:setup}(A), grounded BLG is placed in proximity to a two dimensional Mott insulator multilayer. Because of the mismatch in surface work functions, carriers can transfer between the two subsystems: electrons from BLG may occupy the unfilled upper Hubbard band of the Mott insulator, leaving holes in BLG, or the opposite transfer process may occur depending on the chemical potential alignment, as illustrated in Fig.~\ref{fig:setup}(B). The heavy MI carriers strongly suppress the center-of-mass motion of electron--hole pairs, thereby favoring spontaneous crystallization. Related graphene--correlated insulator heterostructures have been considered in a recent work~\cite{lu_synergistic_2023}, where an electronic crystal in the insulating layer primarily acts as an electrostatic superlattice potential that reconstructs the monolayer graphene bands. Here, by contrast, we focus on the charge-neutral transfer regime, in which the density of transferred carriers in BLG matches that of opposite ones in the MI layer.

In practice, the two electrostatic gates can independently tune the carrier density and the displacement field, which we therefore treat as separate control parameters. The relative strength of Coulomb interactions in each layer is characterized by the two-dimensional Wigner--Seitz parameter $r_s^\alpha=1/(a_{B,\alpha}^{*}\sqrt{\pi n})$, where $\alpha=f,c$ labels the MI and BLG carriers, respectively. $n$ is the transferred carrier density and $a_{B,\alpha}^{*}=4\pi\varepsilon_0
\varepsilon\hbar^2/(m_\alpha^{*}e^2)$ is the corresponding effective Bohr radius. Equivalently, $r_s^\alpha$ measures the characteristic Coulomb interaction energy relative to the kinetic-energy scale. Due to the strong mass asymmetry between the two layers and the low transferred electron--hole density, the heavy MI carriers lie in the large-$r_s^f$ regime and are therefore nearly localized and semiclassical, whereas the BLG carriers remain relatively itinerant, with $r_s^c=\mathcal{O}(1)$. The localized MI carriers consequently form a crystalline charge configuration, while the itinerant BLG carriers bind to them through interlayer Coulomb attraction. The resulting ordered state constitutes a self-generated interlayer electron--hole superlattice.

The key result of this work is that this two-component electron crystal does not need to follow the conventional triangular ordering. In the classical dilute limit, the system approaches a dipolar Wigner crystal \cite{15skinnerprb,skinner16Exciton}, for which Coulomb repulsion favors triangular packing. Away from this limit, however, the itinerant BLG carriers can reorganize into more delocalized charge textures. The associated kinetic and exchange energy gains can overcome the classical electrostatic preference for triangular order and stabilize honeycomb or kagome electron crystals. Depending on displacement field and carrier density, these nonclassical crystals can carry quantum anomalous Hall or quantum spin Hall responses. Our results therefore identify a route to topological electron crystallization in charge-transfer heterostructures, where interlayer attraction, intralayer repulsion and band topology cooperate to stabilize charge orders beyond the conventional Wigner-crystal paradigm \cite{Recher_wigner_blg_2017, joy2023wigner,wanger_wc_blg_2025, skinner_chiral_wigner_2025,Kim_wc_tunneling_2025}. Throughout this work, ``electron crystal'' denotes a two-component (electron and hole) charge-neutral crystalline state as an umbrella term. Its localized classical limit is the dipolar Wigner crystal (DWC), a crystalline order of interlayer excitons; the topologically nontrivial members are labeled by both geometry and response, such as honeycomb quantum spin Hall crystal (QSHC) and kagome quantum anomalous Hall crystal (QAHC).

\section{Model}
\label{sec:model}

We begin by establishing an effective electron--hole bilayer Hamiltonian for the BLG--MI system in the charge-neutral transfer regime,
\begin{equation}
\begin{aligned}
H
&=
\sum_{\bk}
h_c(\bk)\,
c^\dagger_{\bk} c_{\bk}
+
\frac{1}{2}
\sum_{\bk\bk'\bq}
v(\bq)\,
c^\dagger_{\bk+\bq}
c^\dagger_{\bk'-\bq}
c_{\bk'}
c_{\bk}
\\
&\quad
+
\sum_{\bk}
h_f(\bk)\,
f^\dagger_{\bk} f_{\bk}
+
\frac{1}{2}
\sum_{\bk\bk'\bq}
v(\bq)\,
f^\dagger_{\bk+\bq}
f^\dagger_{\bk'-\bq}
f_{\bk'}
f_{\bk}
\\
&\quad
-
\sum_{\bk\bk'\bq}
v'(\bq)\,
c^\dagger_{\bk+\bq}
f^\dagger_{\bk'-\bq}
f_{\bk'}
c_{\bk}.
\end{aligned}
\label{eq:H}
\end{equation}
Here $c^\dagger_{\bk}$ creates a BLG electron and $f^\dagger_{\bk}$ creates a hole in the MI layer. The BLG electrons are described by $h_c(\bk)$, a four-band $\bm{k}\cdot\bm{p}$ Hamiltonian near the $K$ and $K'$ valleys for each spin, including both layer and sublattice degrees of freedom. The $f$-holes originate from the nearly flat lower Hubbard band of the MI surface. The intralayer and interlayer Coulomb interactions are $v(\bq) = e^2/(2S\varepsilon_0 \varepsilon q)$ and $v'(\bq) = e^2 e^{-qd_0}/(2S\varepsilon_0 \varepsilon q)$,
where $d_0$ is the interlayer distance, $S$ is the real-space area, and $q=|\bq|$. Equation~\eqref{eq:H} has a $\mathrm{U}(1)_c\times\mathrm{U}(1)_f$ symmetry, corresponding to independent charge conservation in the two layers. Similar electron--hole bilayer Hamiltonians have been used previously to study excitonic and electron--hole correlated phases \cite{zhu_exciton_prl_1995,franz_exciton_2009,pikulin_exciton_qsh_2014,zhu_exciton_sa_2019,zeng_eh_prb_2020,shao_exciton_prb_204,24jihangprb,25jihangeh}.

Although both charge-transfer directions are in principle possible, we focus on the case in which electrons are transferred from the MI to BLG, leaving holes in the MI layer. We further impose charge-neutral transfer, so that the electron density in BLG equals the hole density in the MI layer, denoted by $n$. Experimentally, the direction and magnitude of charge transfer are mainly determined by the work-function mismatch and can be further tuned electrostatically, while the gate-controlled displacement field $D$ sets the layer potential difference $\Delta$ in BLG. Under our convention, the sign of $D$ determines which BLG layer predominantly hosts the low-energy electron-doped states, see Fig.~\ref{fig:setup}(C). The opposite charge-transfer direction leads to analogous physics, see Note~\ref{sec:S-transfer} of the Supplemental Material.

In the dilute regime considered in this work, $n\sim 10^{11}~{\rm cm}^{-2}$, the two layers lie in very different interaction regimes because of the strong mass asymmetry. For the BLG carriers, a typical effective mass $m_c\simeq 0.033\,m_e$ \cite{li16effectivemass} gives $r_s^c=\mathcal{O}(1)$, so the $c$-electrons remain relatively itinerant. By contrast, the $f$-holes originate from a nearly flat MI band. We therefore take $h_f(\bk)=0$. In this large-$r_s^f$ limit, the $f$-hole dynamics freeze into a semiclassical Wigner crystal like charge order \cite{bonsall77static,goldoni96WCphonon, 15skinnerprb,skinner16Exciton,ambuj25phonon}. The resulting problem is an extended Falicov-Kimball model \cite{freericks_fk_dmft_2003}: interacting BLG electrons move in the electrostatic potential generated by a static configuration of localized $f$-holes, while the full long-range Coulomb energy of the combined electron--hole bilayer is retained. This approximation focuses on the layer $\mathrm{U}(1)$ preserving charge ordered sector and does not include interlayer coherent (excitonic) order.

For a given candidate charge order $\mathcal C$ in the MI layer, the $f$-hole density is approximated as
\begin{equation}
\rho_f^{\mathcal C}(\mathbf r)
=
\sum_{i}
\sum_{\bm{\tau}\in\mathcal B_{\mathcal C}}
\delta(\mathbf r-\mathbf R_i-\bm{\tau}),
\end{equation}
where $\mathbf R$ runs over the Bravais lattice vectors of the $f$-hole crystal and $\mathcal B_{\mathcal C}$ denotes the set of basis positions in one unit cell. We restrict to charge configurations preserving $C_{3z}$ rotational symmetry. For a crystal configuration with $N_b$ $f$-hole sites per unit cell, the unit-cell area is therefore $\Omega_{\mathcal C}=N_b/n$. We use a triangular Bravais lattice for all candidate states, with $\Omega_{\mathcal C}=\sqrt{3}L_{\mathcal C}^2/2$, so that
$L_{\mathcal C}=\sqrt{2N_b/(\sqrt{3}n)}$.
Triangular, honeycomb, and kagome orders are implemented by choosing $N_b=1,2,3$, respectively, together with the corresponding basis positions within the triangular Bravais cell.

We solve the BLG $c$-electron problem self-consistently at the Hartree Fock level to obtain the density matrix $\rho_c^\mathcal{C}$ for each fixed $f$-hole configuration. The static $f$-hole crystal generates an internally produced periodic interlayer potential for the BLG electrons. In the mini Brillouin zone representation, this term is
\begin{equation}
H_{\rm inter}^{\mathcal C}
=
\sum_{\bar{\bk}}
\sum_{\bG,\bQ}
U_{\mathcal C}(\bQ)\,
c^\dagger_{\bar{\bk},\bG}
c_{\bar{\bk},\bG+\bQ},
\end{equation}
with
\begin{equation}
\begin{aligned}
U_{\mathcal C}(\bQ)
&=
-
\frac{e^2}{2\varepsilon_0\varepsilon\Omega_{\mathcal C}}
\frac{e^{-|\bQ|d_0}}{|\bQ|}
S_{\mathcal C}(\bQ),
\\
S_{\mathcal C}(\bQ)
&=
\sum_{\bm{\tau}\in\mathcal B_{\mathcal C}}
e^{-i\bQ\cdot\bm{\tau}}.
\end{aligned}
\end{equation}
Here $\bar{\bk}$ lies in the mini Brillouin zone, $\bG$ and $\bQ$ are reciprocal lattice vectors of the $f$-hole crystal, and $S_{\mathcal C}(\bQ)$ is the structure factor of the chosen charge order. This periodic potential folds the BLG bands into the mini Brillouin zone and can open gaps or modify the miniband topology. Upon obtaining the self-consistent density matrix for a given BLG crystal configuration $\mathcal C$, we calculate the total energy as
\begin{equation}
E_{\rm T}^{\mathcal C}
=
E_{\rm kin}^{c}[
\rho_c^{\mathcal C}
]
+
E_{\rm F}^{c}[
\rho_c^{\mathcal C}
]
+
E_{\rm H}
[
\rho_c^{\mathcal C},
\rho_f^{\mathcal C}
],
\label{eq:Etot}
\end{equation}
where $E_{\rm kin}^{c}$ and $E_{\rm F}^{c}$ are the kinetic and exchange energies of the BLG $c$-electrons, and $E_{\rm H}$ is the classical electrostatic energy of the full electron--hole bilayer. Explicitly,
\begin{widetext}
\begin{equation}
\begin{aligned}
E_{\rm H}
&=
\frac{1}{2}
\int d\mathbf r\,d\mathbf r'\,
\frac{e^2}
{4\pi\varepsilon_0\varepsilon|\mathbf r-\mathbf r'|}
\left[
\rho_c(\mathbf r)\rho_c(\mathbf r')
+
\rho_f(\mathbf r)\rho_f(\mathbf r')
\right]
-
\int d\mathbf r\,d\mathbf r'\,
\frac{e^2}
{4\pi\varepsilon_0\varepsilon
\sqrt{|\mathbf r-\mathbf r'|^2+d_0^2}}
\rho_c(\mathbf r)
\rho_f(\mathbf r'). \label{eq:en}
\end{aligned}
\end{equation}
\end{widetext}
This Hartree energy is evaluated for the charge-neutral bilayer configuration using a single Ewald summation, because each layer separately carries a nonzero net charge and its long-range Coulomb energy diverges. Only the combined charge-neutral electron--hole bilayer has a well defined electrostatic energy. The main approximation in the total energy evaluation is that we neglect the quantum corrections associated with the finite mass of the $f$-holes. The leading correction is the zero-point motion of the $f$-holes, whose energy scales as $\mathcal{O}[(r_s^f)^{-3/2}]$ and is therefore smaller than the classical Hartree energy $\mathcal{O}[(r_s^f)^{-1}]$ in the large-$r_s^f$ regime. The same zero-point motion also gives the localized $f$-holes a finite spatial width, which modifies the superlattice potential felt by the itinerant $c$-electrons and yields an additional energy correction on the order of $\mathcal{O}[(r_s^f)^{-3/2}]$. Corrections from the finite $f$-band dispersion and exchange interaction between localized $f$-holes are further suppressed in the flat-band, large-$r_s^f$ limit.

\begin{table}[t]
\caption{\label{tab:chern}
Valley Chern numbers of the active BLG miniband induced by different $f$-carrier superlattice geometries in the MI layer. The triple $(D,\mathrm{MI},\mathrm{BLG})$ specifies the sign of the displacement field and the carrier types left in the MI and BLG layers after charge transfer, where $e$ and $h$ denote electrons and holes, respectively. For electron-doped BLG, $C_c$ is the valley Chern number of the first conduction miniband; for hole-doped BLG, $C_v$ is that of the first valence miniband. The triangular $f$-hole lattice always yields a topologically trivial active miniband, whereas the honeycomb and kagome lattices generate topological minibands with $|C| = 1$, the sign of which is set by the displacement-field convention and the electron/hole channel.}
\begin{ruledtabular}
\begin{tabular}{lccc}
$(D,\mathrm{MI},\mathrm{BLG})$ & Triangular & Honeycomb & Kagome \\
\colrule
Case 1: $(D>0,\,h,\,e)$ & $C_c = 0$  & $C_c = -1$ & $C_c = -1$ \\
Case 2: $(D<0,\,h,\,e)$ & $C_c = 0$  & $C_c = +1$ & $C_c = +1$ \\
Case 3: $(D>0,\,e,\,h)$ & $C_v = 0$  & $C_v = +1$ & $C_v = +1$ \\
Case 4: $(D<0,\,e,\,h)$ & $C_v = 0$  & $C_v = -1$ & $C_v = -1$ \\
\end{tabular}
\end{ruledtabular}
\end{table}

In this semiclassical approximation, the $f$-hole Wigner crystal provides a static charge-order background for the itinerant BLG electrons. In the parameter regime considered here, the interlayer Coulomb potential acts as a periodic electrostatic perturbation to the BLG bands, rather than driving the system into the strong-coupling limit of tightly localized electron--hole dipoles. The resulting miniband topology depends on both the charge-transfer direction and the sign of the displacement field. We summarize in Table~\ref{tab:chern} the valley Chern number of the active BLG miniband for the different displacement-field and charge-transfer conventions. In the main text, we focus on the regime where electrons are transferred from the MI to BLG, leaving holes in the MI layer, corresponding to Cases 1 and 2.

\section{Dipolar Wigner crystal in the classical limit}
\label{sec:dwc}

\begin{figure}[t]
\centering
\includegraphics[width=\columnwidth]{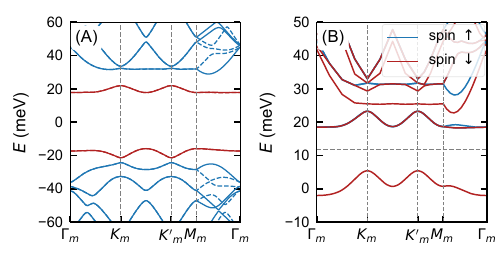}
\caption{\label{fig:minibands}
Bilayer graphene minibands.
(A) BLG energy bands subject to the superlattice potential of a triangular arrangement of $f$-holes in the MI layer, for $\Delta = 40$~meV and $L = 30$~nm. The transferred electrons dope the first conduction band; the potential difference is $\Delta = |e|Dd/\varepsilon$ with $d = 3.35$~\AA.
(B) Hartree Fock mean-field bands for BLG at the filling of one electron per superlattice unit cell for the same parameters as used in (A). The chemical potential is labeled as the gray dashed line.}
\end{figure}

We first consider the extreme dilute limit of the electron--hole density, where the natural reference state is the triangular dipolar Wigner crystal. In this regime, the electrostatic energy dominates and favors a triangular arrangement of dipoles. We therefore take the holes in the MI layer to form a triangular lattice.

The periodic interlayer Coulomb potential generated by this hole crystal reconstructs the BLG spectrum, as shown in Fig.~\ref{fig:minibands}(A). In the presence of a displacement field, the superlattice potential folds and flattens the original BLG bands, producing narrow conduction and valence minibands for each spin and valley flavor. The first conduction miniband is topologically trivial because electrons reside near the potential minima and thus inherit a triangular pattern \cite{zeng_gate-tunable_2024}.

\begin{figure*}[t]
\centering
\includegraphics[width=0.95\textwidth]{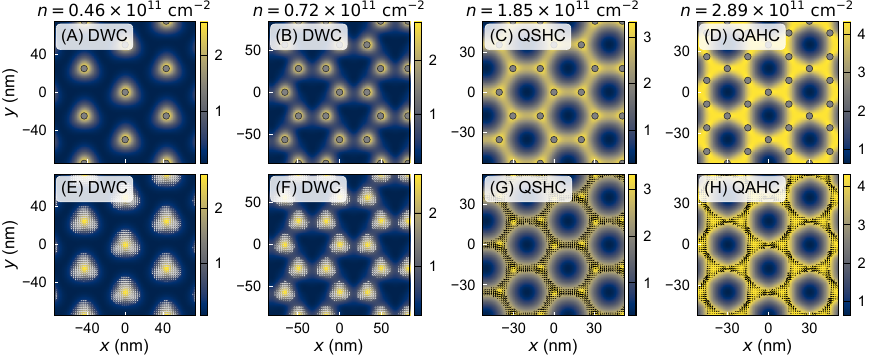}
\caption{\label{fig:density}
Ground-state charge density and current distributions.
(A--D) Real-space charge density $\rho_c(\mathbf{r})\,\Omega_\mathcal{C}$ on the BLG layer obtained from self-consistent Hartree Fock calculations at $\Delta = 40$~meV; gray circles mark the localized holes in the MI layer. With increasing carrier density $n$, the system undergoes a density-driven quantum melting from (A) a triangular dipolar Wigner crystal (triangular DWC) into (B) a honeycomb dipolar Wigner crystal (honeycomb DWC) and (C) a honeycomb quantum spin Hall crystal (honeycomb QSHC), followed by (D) a kagome quantum anomalous Hall crystal (kagome QAHC), demonstrating a progressive delocalization of the BLG electrons from site-centered dipoles into topological charge orders.
(E--H) Calculated charge current $\bm{j}_c(\mathbf{r})$ on the BLG side corresponding to the phases in (A)--(D), respectively.}
\end{figure*}

At one transferred electron per superlattice unit cell, the first conduction miniband is quarter filled before interactions. The narrow miniband width enhances exchange and drives spontaneous flavor polarization. Self-consistent Hartree Fock calculations indeed yield a spin--valley polarized correlated insulator with total Chern number \(C=0\) [Fig.~\ref{fig:minibands}(B)]. In this state, the BLG electron density develops maxima directly opposite to the holes in the MI layer, forming a triangular dipolar Wigner crystal (triangular DWC). A representative real-space density profile is shown in Fig.~\ref{fig:density}(A). This tendency can be understood using a classical electrostatic analogue of a charged layer separated by a distance $d_0$ from a grounded metallic plate: the induced opposite charge accumulates directly beneath the external charges to minimize the electrostatic energy, yielding the conventional triangular DWC \cite{15skinnerprb, Topping1927-wm}. This classical parallel-plate capacitor model is solved using the image charge method, see Note~\ref{sec:S-classical} of the Supplemental Material. We use this self-consistent triangular DWC as the reference state for comparison with the honeycomb and kagome electron crystals discussed below.

\section{Competing charge orders and phase diagram}
\label{sec:phase}

We next ask whether alternative \(C_3\)-symmetric charge orders can become competitive away from the extreme dilute limit. To address this question, we consider honeycomb and kagome arrangements of the localized \(f\)-holes. These candidate lattices are constructed at fixed total electron--hole pair density, enabling a controlled comparison among distinct charge-ordering patterns.

The interlayer Coulomb potential experienced by the itinerant $c$-electrons is determined by the real-space geometry of the $f$-hole lattice. Different from the triangular case, honeycomb and kagome arrangements invert the effective potential landscape: the potential minima form honeycomb or kagome networks, while the maxima lie on triangular sites. As a result, the first conduction miniband becomes topological, with valley Chern number $|C|=1$ and the sign of valley Chern number is determined by displacement field $D$ (Table~\ref{tab:chern}). This sign reversal demonstrates that the nontrivial topology of the miniband arises mainly from the interplay between the superlattice potential and the Berry curvature inherited from bilayer graphene, see Note~\ref{sec:S-topo} of the Supplemental Material for a detailed perturbation theory discussion.

We then perform self-consistent Hartree Fock calculations for these candidate lattices, allowing for spontaneous flavor symmetry breaking. Representative converged real-space charge densities $\rho_c(\mathbf{r})$ of the BLG electrons are shown in Fig.~\ref{fig:density}, together with the corresponding $f$-hole positions in the MI layer. Under honeycomb and kagome $f$-hole backgrounds, the BLG electrons reorganize into matching nontriangular charge textures, in clear contrast to the triangular reference state. In the honeycomb case, a self-consistent mean-field solution is a honeycomb quantum spin Hall crystal (honeycomb QSHC), in which two electrons occupy opposite spin--valley flavors ($K\uparrow$, $K'\downarrow$). When only long-range Coulomb interactions are retained within the BLG subsystem, the emergent SU(2)$\times$SU(2) symmetry leads to a degeneracy among quantum spin Hall, quantum valley Hall, and spin-valley-locked states. In the kagome case, three electrons fill three of the four spin-valley flavors, yielding a kagome quantum anomalous Hall crystal (kagome QAHC). These self-consistent topological electron crystals are therefore both structurally and topologically distinct from the triangular DWC. The remaining question is whether they can also overcome the triangular state energetically.

\begin{figure*}[t]
\centering
\includegraphics[width=0.86\textwidth]{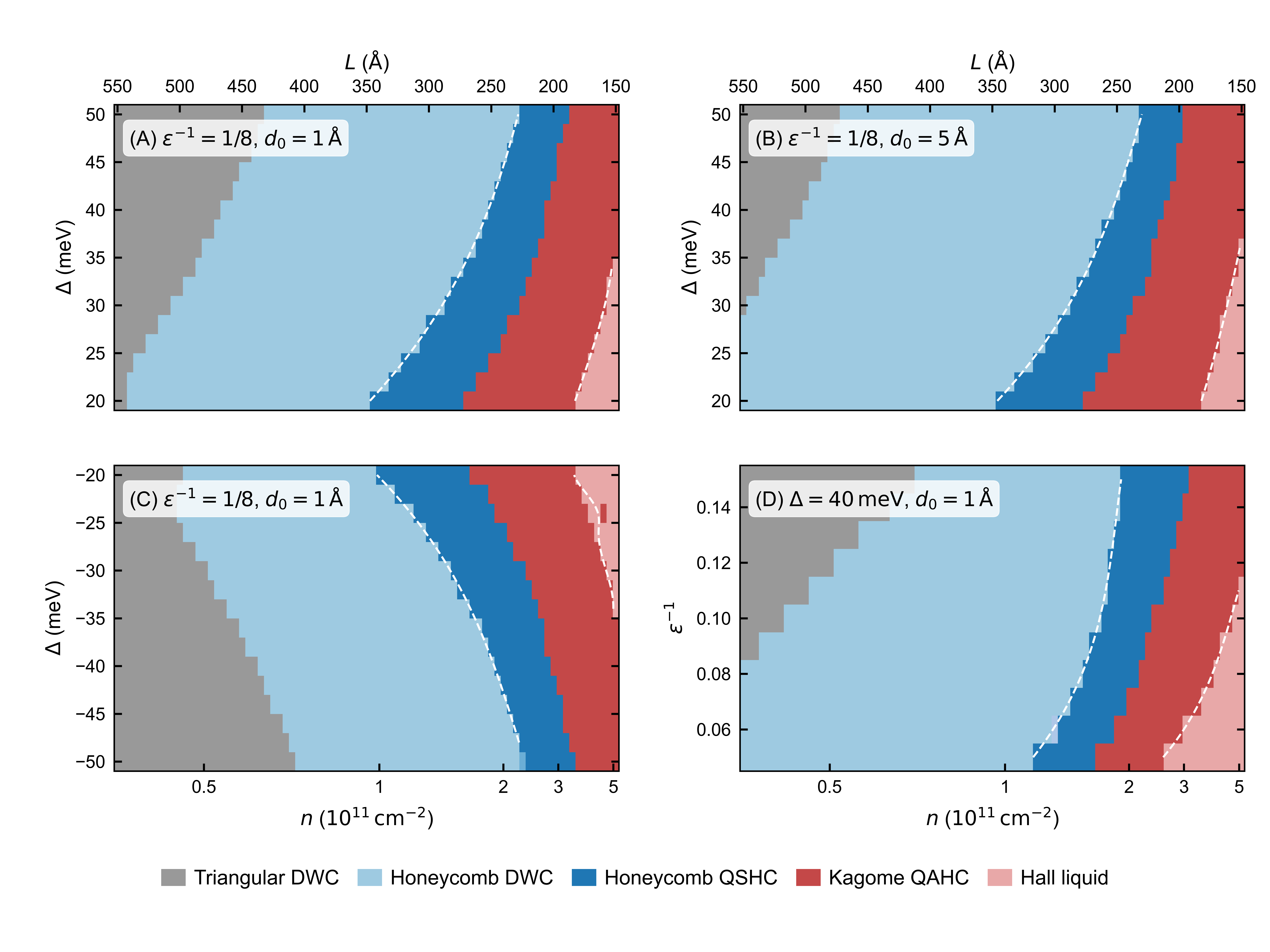}
\caption{\label{fig:phase}
Phase diagrams.
(A--D) Calculated phase diagrams comparing total energy as a function of the electron--hole pair density $n$. (A), (B) positive potential difference (positive displacement field, $\Delta > 0$) with different interlayer distance $d_0$, (C) negative potential difference (negative displacement field, $\Delta < 0$), and (D) effective interaction strength $\varepsilon^{-1}$. Colored regions represent the ground-state phases, labeled by their lattice geometry and topological invariants: kagome quantum anomalous Hall crystal (kagome QAHC, $|C| = 1$), honeycomb quantum spin Hall crystal (honeycomb QSHC, $C_s = 1$), and dipolar Wigner crystal (DWC). White dashed lines indicate the phase boundaries of the topological electron crystals. $L$ denotes the lattice constant for the triangular lattice case.}
\end{figure*}

We compare the total energies for each electron--hole pair [Eq.~\eqref{eq:Etot}] of the triangular, honeycomb, and kagome charge-ordered states to determine the ground state phase diagram. The phase diagram [Fig.~\ref{fig:phase}] reveals a systematic evolution of the ground state with electron--hole density $n$ and the layer potential difference $\Delta$. In the dilute limit, the system forms a triangular DWC. As the density increases, the system evolves through honeycomb DWC, honeycomb QSHC, and kagome QAHC phases, and in part of the phase diagram eventually enters a Hall liquid regime at the highest densities. Here the Hall liquid denotes a translationally invariant Hartree Fock state with a nonzero Hall response but without the electron crystal charge modulation.

\begin{figure*}[t]
\centering
\includegraphics[width=0.95\textwidth]{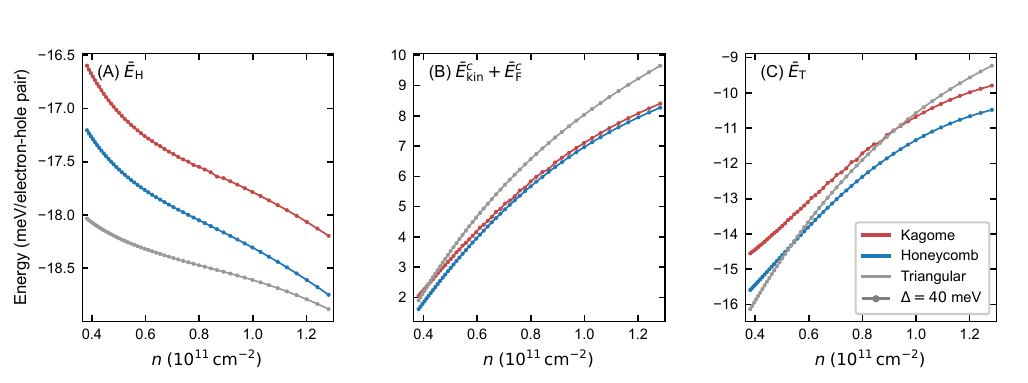}
\caption{\label{fig:energy}
Energy decomposition.
(A) Hartree energy per electron--hole pair, \(\bar{E}_{\mathrm{H}}\), as a function of density \(n\) for different candidate lattice geometries. The triangular lattice has the lowest Hartree energy throughout the density range. Parameters are fixed at \(\varepsilon = 8\) and \(\Delta = 40~\mathrm{meV}\). The macroscopic capacitor energy has been subtracted.
(B) Combined kinetic and Fock (exchange) energies, representing the quantum contribution to the total energy.
(C) Total energy per electron--hole pair $\bar{E}_\mathrm{T} = \bar{E}_\mathrm{H} + \bar{E}_\mathrm{kin}^c + \bar{E}_\mathrm{F}^c$. The competition between the classical Hartree preference and quantum delocalization leads to a crossing of the energy curves, stabilizing the honeycomb electron crystal over the triangular DWC as $n$ increases.}
\end{figure*}

This evolution reflects a progressive departure from the classical dipolar limit. In the triangular DWC, the BLG electrons remain largely localized above the holes in the MI layer, forming an ionic-like charge configuration stabilized primarily by classical electrostatics. By contrast, in the honeycomb QSHC and kagome QAHC phases, the reconstructed BLG minibands allow the electronic charge distribution $\rho_c(\mathbf{r})$ to spread away from the hole sites and develop bonding patterns, as shown in Figs.~\ref{fig:density}(C) and \ref{fig:density}(D). These states are therefore better viewed as covalent-like electron crystals: delocalization of the BLG electrons lowers the kinetic and exchange energies and drives intermediate ``quantum-melted'' crystals between the site-centered dipole solid and the fully delocalized liquid. We emphasize that the topology discussed here is that of the self-consistent Hartree Fock bands, which does not need to coincide with the topology of the single particle $c$ bands in the static $f$-crystal potential. This distinction is important in the honeycomb case. Although the single particle honeycomb potential favors a topological miniband structure, the low density Hartree Fock solution is a topologically trivial antiferromagnetic honeycomb DWC, where interaction generates an additional gap. With increasing density, the electrons become more itinerant and the honeycomb QSHC emerges through a gap closing transition. In this sense, the trivial antiferromagnetic honeycomb DWC is a precursor to the honeycomb QSHC in the density driven quantum melting process.

This picture is supported by both the parameter scans in Figs.~\ref{fig:phase}(B) and \ref{fig:phase}(C) and the energy decomposition in Fig.~\ref{fig:energy}. The asymmetry with respect to $\Delta$ reflects the layer polarized nature of the BLG minibands, while the expansion of the triangular DWC region with increasing $\varepsilon^{-1}$ confirms its classical electrostatic origin. Consistently, Fig.~\ref{fig:energy} shows that the triangular DWC is favored by its lower Hartree energy, whereas the honeycomb and kagome electron crystals pay a larger Hartree cost but gain kinetic and Fock energies through more delocalized BLG charge textures. The phase boundaries are therefore set by total energy crossings between classical electrostatic localization and quantum delocalization. We have also checked the robustness of the phase diagram against the BLG--MI separation \(d_0\).
As shown in Figs.~\ref{fig:phase}(B) and \ref{fig:phaseSM}, increasing \(d_0\) modifies the phase boundaries quantitatively but preserves the same phase sequence.

\section{Discussion}
\label{sec:discussion}

The BLG-MI heterostructure studied here provides a minimal electron-hole bilayer platform that can mimic a two-dimensional solid, in which a heavy crystalline charge order electrostatically couples to itinerant BLG carriers with non-trivial Berry curvature. In the dilute, classical electrostatic limit, Coulomb interaction favors a simple triangular dipolar Wigner crystal (DWC) of interlayer excitons: each electron--hole pair is tightly bound across the layers. Our central result is that as carrier density increases, this localized-dipole picture breaks down. While the $f$-holes remain sufficiently heavy to form a static background, the internally generated periodic potential reconstructs the BLG minibands, allowing the itinerant electrons to delocalize into bonding charge textures away from the hole sites.

As a result, the crystal geometry is not solely determined by the classical electrostatics of the heavy charges. Instead, it is selected by a competition between the electrostatic cost of distorting the $f$-hole Wigner crystal and the quantum energy gained by the itinerant BLG electrons. When the latter dominates, the energy gain can effectively drag the heavy holes away from their electrostatically preferred triangular arrangement and stabilize honeycomb or kagome charge orders. The resulting phases are intrinsically quantum electron crystals: their bonding in BLG layer is more covalent-like than ionic-like, and in specific density and displacement field regimes the reconstructed BLG minibands acquire nontrivial topology. Furthermore, these phases act as intermediate states bridging the localized triangular DWC and the fully itinerant Hall-liquid limit during a density driven quantum melting process.

We next discuss possible materials platforms. For the semiclassical treatment of the $f$-holes to be valid, the $f$ layer should be deep in the large-$r_s^f$ regime. For the dilute densities considered in this work, $n\approx 10^{11}~{\rm cm}^{-2}$, we estimate $r_s=1/(a_B^*\sqrt{\pi n})$ with $a_B^*$ the effective Bohr radius.
Thus, for BLG with $m_c\simeq 0.033\,m_e$~\cite{li16effectivemass}, $\varepsilon \simeq 5$, one obtains $r_s^c\simeq 2$, so the BLG carriers remain relatively itinerant. By contrast, a heavy or moir\'e-flattened $f$ band with $m_f\gtrsim m_e$ already gives $r_s^f\gg 10$, while masses of order a few $m_e$ place the $f$-holes deep in the semiclassical Wigner-crystal regime.

Two classes of materials are particularly relevant. The first is BLG coupled to a moir\'e TMD layer, such as twisted WS$_2$ \cite{zhang_engineering_2024} or WSe$_2$/WS$_2$ heterostructures \cite{qi2026observationexciton}. In these systems, the relatively large TMD effective mass, together with the additional bandwidth suppression from the moir\'e potential, can strongly enhance the tendency toward charge localization in the $f$ layer \cite{You2026genWC}. The moir\'e potential may also provide an additional pinning mechanism that stabilizes the crystalline charge order. The second class consists of BLG coupled to layered Mott insulators like CrOCl \cite{yang_unconventional_2023}, 1T-TaS$_2$ \cite{tilak2024proximity}, and Nb$_3$X$_8$ $(X={\rm Cl,Br,I})$ \cite{zhang2023mottness,gao2023MI,Yang2025-kw,2506.21837}. Recent theoretical calculations and transport measurements in these materials have reported remarkably large carrier effective masses, reaching several times the free electron mass $m_e$. Furthermore, the graphene-Nb$_3$Cl$_8$ heterostructure exhibits an anomalous screening effect in transport measurement \cite{Yang2025-kw} (often referred to as the ``gate-does-not-work'' feature \cite{Skinner2023}), which suggests that the charge transfer process can be continuously tuned by external gates. More broadly, our mechanism should apply to itinerant layers with nontrivial band geometry coupled electrostatically to a heavy charge-ordered layer. Possible itinerant platforms include rhombohedral graphene multilayers \cite{dong_anomalous_2024,soejima_anomalous_2024,dong_theory_2024, dong_stability_2024, zhou_fractional_2024, zhou_new_2025, desrochers_elastic_2025, guo_correlation_2025, kwan_rmg_prb_2025, yu_rmg_ed_prb_2025,
miao25electron, bab_trashcan_2025}, Rashba/Ising-type two-dimensional electron gases \cite{yang_topological_2024,pi_moire_ising}, and BHZ-type quantum wells \cite{eng_cd3as2, tan_designing_2024, miao_artificial_2025,yang_engineering_2025,liu_ideal_heavyf_2025, efficient_predict_25}.

Several questions remain beyond the present mean-field treatment. Relaxing the heavy $f$-hole limit would introduce quantum fluctuations of the localized charges and may drive competing excitonic \cite{25jihangeh}, supersolid \cite{sarma06supersolid}, or superconducting instabilities \cite{2601.07729}. In particular, allowing interlayer coherence could reveal exciton supersolid phases in which excitonic order coexists with the density modulation of the electron crystal \cite{sarma06supersolid}. Away from the charge-neutral transfer regime, excess carriers may bind to the dipoles, giving rise to doped electron crystals or trionic phases \cite{fu_trion_2024}. Finally, the reconstructed topological minibands may provide a route to correlated states at noninteger fillings, including fractional Chern insulating phases \cite{levin2009fractional,qi2011generic,kang2024evidence,zhang2024vortex,jian2025minimal,zou_valley_order}.

\textit{Note Added.} After completion of this work, a recent Letter reported charge transfer induced heavy Wigner crystal phases in a related setting \cite{heavyf_wigner_prl_2025}. We also note recent preprints discussing charge transfer induced correlated phases in CrOCl--rhombohedral graphene heterostructures \cite{shi2026correlated} and superconducting instabilities in related electron--hole bilayer systems \cite{2601.07729}.

\begin{acknowledgments}
W. Miao thanks F. Desrochers, C.X. Liu and K.F. Luo for inspiring discussion. L. Wang acknowledges the support from National Key Projects for Research and Development of China (Grant No. 2022YFA1204700 \& 2021YFA1400400), National Natural Science Foundation of China (Grant No. 12525403 \& 12550404) and Natural Science Foundation of Jiangsu Province (Grant No. BK20253027 \& BK2023300). X.Y. Song is supported by the Early Career Scheme of Hong Kong Research Grant Council (Grant No. 26309524). X. Dai is supported by a fellowship award from Hong Kong Research Grant Council (Project No. SRFS2324-6S01) and New Cornerstone Science Foundation.
\end{acknowledgments}

\onecolumngrid

\appendix

\section{Continuum \texorpdfstring{$\bm{k}\cdot\bm{p}$}{k.p} model for bilayer graphene}
\label{app:kp}

All calculations are performed within the semiclassical Falicov--Kimball framework introduced in the main text: the $f$-holes form a static $C_{3z}$-symmetric charge order in the MI layer [Eq.~\eqref{eq:H} with $h_f(\bk)=0$], and the itinerant BLG $c$-electrons are treated self-consistently at the Hartree Fock level in the periodic interlayer potential $U_{\mathcal C}(\bQ)$ that this background generates. Below we specify the continuum BLG Hamiltonian $h_c(\bk)$, the self-consistent Hartree Fock procedure, and the Ewald scheme used to evaluate the total electrostatic energy of the charge-neutral electron--hole bilayer. Throughout, $\varepsilon$ is the relative dielectric constant, $d=3.35$~\AA\ is the interlayer spacing within BLG, and $d_0$ is the BLG--MI separation entering the interlayer interaction $v'(\bq)$ of Eq.~\eqref{eq:H}.

The single-particle term $h_c(\bk)$ in Eq.~\eqref{eq:H} is the four-band $\bm{k}\cdot\bm{p}$ Hamiltonian of Bernal-stacked (AB) bilayer graphene, expanded near the $K$ and $K'$ valleys. In the sublattice--layer basis $(A_1,B_1,A_2,B_2)$, where $A/B$ label the two sublattices and $1/2$ the bottom and top graphene layers, the Hamiltonian for valley $\xi=\pm1$ (corresponding to $K$ and $K'$) and each spin reads
\begin{equation}
h_c(\bk)\equiv H_\xi(\bk) =
\begin{pmatrix}
-\Delta/2 & v_0 \pi_\xi^{\dagger} & -v_4 \pi_\xi^{\dagger} & v_3 \pi_\xi \\
v_0 \pi_\xi & -\Delta/2 + \Delta' & \gamma_1 & -v_4 \pi_\xi^{\dagger} \\
-v_4 \pi_\xi & \gamma_1 & \Delta/2 + \Delta' & v_0 \pi_\xi^{\dagger} \\
v_3 \pi_\xi^{\dagger} & -v_4 \pi_\xi & v_0 \pi_\xi & \Delta/2
\end{pmatrix},
\label{eq:hc}
\end{equation}
with $\pi_\xi = \xi k_x + i k_y$. The parameters are the interlayer dimer hopping between $(B_1,A_2)$ sites $\gamma_1=0.40$~eV, which sets the overall bilayer band structure; the intralayer Dirac velocity $v_0=6.51$~eV\,\AA; the trigonal-warping velocity $v_3=(\sqrt{3}/2)a\gamma_3$ with $\gamma_3=0.315$~eV; the skew interlayer velocity $v_4=(\sqrt{3}/2)a\gamma_4$ with $\gamma_4=0.044$~eV, responsible for electron--hole asymmetry; and a small onsite sublattice-energy difference $\Delta'=0.018$~eV. The lattice constant is $a=2.46$~\AA. The interlayer potential difference $\Delta=|e|Dd/\varepsilon$ is controlled by the gate-induced displacement field $D$ and opens a tunable gap at charge neutrality while layer-polarizing the low-energy states. This Hamiltonian is included for both spin and both valley flavors; the heavy $f$-holes are treated as a static classical background, $h_f(\bk)=0$.

\section{Self-consistent mean field treatment of the BLG electrons}
\label{app:hf}

For each fixed $f$-hole configuration $\mathcal C$, the single-particle Hamiltonian for BLG can be taken as
\begin{equation}
    H_0(\bkbar)=h_c\!\left(\bk\to-i\partial_{\mathbf r}\right)+U_{\mathcal C}(\mathbf r),
\end{equation}
which combines the continuum model of Eq.~\eqref{eq:hc} with the periodic interlayer potential $U_{\mathcal C}(\bQ)$, both expressed in the mini Brillouin zone (mBZ) basis. We retain the long-range component of the Coulomb interaction and write
\begin{equation}
H_{\mathrm{int}}=\frac{1}{2}\sum_{\bq} v(\bq)\,{:}\,\hat{\rho}_{\bq}\hat{\rho}_{-\bq}\,{:}\,,
\qquad
\hat{\rho}_{\bq}=\sum_{\bk}\sum_{s\xi\alpha}c^{\dagger}_{\bk+\bq,\,s\xi\alpha}\,c_{\bk,\,s\xi\alpha},
\end{equation}
where $s$ is the spin index, $\xi$ the valley index, and $\alpha$ the joint sublattice--layer index. The intralayer interaction within BLG is
\begin{equation}
v(\bq)=\frac{e^{2}}{2S\,\varepsilon_0\varepsilon\, q},
\end{equation}
with $S$ the system area. Projecting onto the active miniband subspace gives
\begin{equation}
\begin{aligned}
    H_{\mathrm{int}} =
\frac{1}{2N_{\bkbar}}
\sum_{\bkbar,\bkbar',\bar{\bq}}
\sum_{\xi\xi'}\sum_{ss'}\sum_{nm n'm'}
&\left( \sum_{\bQ} v(\bQ+\bar{\bq})\,
\lambda^{\xi\xi' ss'}_{nm,n'm'}(\bkbar,\bkbar',\bar{\bq},\bQ) \right)\\
&\times
d^\dagger_{s\xi n}(\bkbar+\bar{\bq})
d^\dagger_{s'\xi'n'}(\bkbar'-\bar{\bq})
d_{s'\xi'm'}(\bkbar')\,
d_{s\xi m}(\bkbar),
\end{aligned}
\end{equation}
where $\bkbar,\bkbar',\bar{\bq}$ are momenta in the mBZ generated by the band folding, $N_{\bkbar}$ is the number of mBZ grid points, and $\bG,\bQ$ are reciprocal lattice vectors of the $f$-hole crystal. The projected form factor is
\begin{equation}
\lambda^{\xi \xi' s s'}_{nm,n'm'}(\bkbar,\bkbar',\bar{\bq},\bQ)
= \sum_{\alpha\alpha'\bG \bG'}
C^{*}_{\xi \alpha s \bG+\bQ,n}(\bkbar+\bar{\bq})\,
C^{*}_{\xi' \alpha's' \bG'-\bQ,n'}(\bkbar'-\bar{\bq})\,
C_{\xi' \alpha' s'\bG',m'}(\bkbar')\,
C_{\xi \alpha s\bG,m}(\bkbar),
\end{equation}
where the coefficients $C$ relate the miniband operators to the plane-wave basis through
\begin{equation}
    d_{s\xi n}^\dagger(\bkbar) = \sum_{\alpha \bG} C_{\xi \alpha s \bG, n}(\bkbar)\, c^\dagger_{\xi \alpha s}(\bkbar+\bG).
\end{equation}
We solve the mean-field problem in the Green's-function (Dyson) framework, using composite band indices $\eta=(s,\xi,n)$. With the single-particle density matrix $\rho_{\eta\eta'}(\bkbar)=\langle c^{\dagger}_{\bkbar,\eta'}c_{\bkbar,\eta}\rangle$, the static Hartree and Fock self-energies are
\begin{align}
[\Sigma^{\rm H}(\bkbar)]_{\eta\eta'} &= \frac{1}{N_{\bkbar}}\sum_{\bkbar',\zeta,\zeta',\bQ} v(\bQ)\,\Lambda^{\rm H}_{\eta\eta';\,\zeta\zeta'}(\bkbar,\bkbar';\bQ)\,\rho_{\zeta'\zeta}(\bkbar'), \\
[\Sigma^{\rm F}(\bkbar)]_{\eta\eta'} &= -\frac{1}{N_{\bkbar}}\sum_{\bkbar',\zeta,\zeta',\bQ} v(\bkbar'-\bkbar+\bQ)\,\Lambda^{\rm F}_{\eta\zeta;\,\zeta'\eta'}(\bkbar,\bkbar';\bQ)\,\rho_{\zeta'\zeta}(\bkbar'),
\end{align}
where the tensors are the projected form factors evaluated at zero ($\Lambda^{\rm H}$) and finite ($\Lambda^{\rm F}$) transferred momentum,
\begin{equation}
\Lambda^{\rm H}_{\eta\eta';\zeta\zeta'}(\bkbar,\bkbar';\bQ)=\lambda^{\xi\xi'}_{nm,n'm'}(\bkbar,\bkbar',\bar{\bq}=0,\bQ),
\qquad
\Lambda^{\rm F}_{\eta\zeta;\zeta'\eta'}(\bkbar,\bkbar';\bQ)=\lambda^{\xi\xi'}_{nm,n'm'}(\bkbar,\bkbar',\bar{\bq}=\bkbar'-\bkbar,\bQ),
\end{equation}
with $\eta=(s,\xi,n)$, $\eta'=(s,\xi,m)$, $\zeta=(s',\xi',m')$, and $\zeta'=(s',\xi',n')$. Starting from a random density matrix that allows spontaneous symmetry breaking among the spin--valley flavors, we build the Dyson Hamiltonian $H_{\mathrm{HF}}=H_0+\Sigma^{\rm H}+\Sigma^{\rm F}$, diagonalize it to update the Green's function and density matrix, and iterate with linear mixing until both $\rho$ and $\Sigma$ converge. At convergence the self-energy defines the BLG quasiparticle spectrum, and the BLG charge density follows from
\begin{equation}
\begin{aligned}
\rho_c(\mathbf{r}) &= \sum_\bQ \rho_c(\bQ)\, e^{i \bQ \cdot\mathbf{r}},\qquad
\rho_c(\bQ)
=
\frac{1}{\Omega_{\mathcal{C}}N_{\bkbar}}
\sum_{\bkbar,\ell} f_{\ell\bkbar}
\sum_{\xi s\alpha\bG}
\mathcal C^{\mathrm{HF}*}_{\xi s\alpha\bG,\ell}(\bkbar)\,
\mathcal C^{\mathrm{HF}}_{\xi s\alpha,\bG+\bQ,\ell}(\bkbar).
\end{aligned}
\end{equation}
Here $\mathcal C^{\mathrm{HF}}_{\xi s\alpha\bG,\ell}(\bkbar)$ is the plane-wave component of the $\ell$-th Hartree Fock eigenstate at momentum $\bkbar$, and $f_{\ell\bkbar}$ is its occupation. The converged solution yields the kinetic and exchange energy density of the BLG $c$-electrons,
\begin{equation}
\mathcal{E}_{\mathrm{kin}}^{c}=\frac{1}{\Omega_\mathcal{C}N_{\bkbar}}\sum_{\bkbar}\mathrm{Tr}\!\left[h_c(\bkbar)\,\rho(\bkbar)\right],
\qquad
\mathcal{E}_{\mathrm F}^{c}=\frac{1}{2\Omega_\mathcal{C} N_{\bkbar}}\sum_{\bkbar}\mathrm{Tr}\!\left[\Sigma^{\rm F}(\bkbar)\,\rho(\bkbar)\right],
\end{equation}
which enter the total energy density decomposition $\mathcal{E}_{\mathrm T}^{\mathcal C}=\mathcal{E}_{\mathrm{kin}}^{c}+\mathcal{E}_{\mathrm F}^{c}+\mathcal{E}_{\mathrm H}$. The classical Hartree energy density $\mathcal{E}_{\mathrm H}$ of the full charge-neutral bilayer which includes the BLG intralayer term is evaluated separately by the Ewald procedure below, so that it is counted once in the total energy. In practice we retain the lowest few conduction minibands for the triangular, honeycomb, and kagome configurations, with a common plane-wave cutoff $\Lambda=3g$, where $g=2\pi\sqrt{2n/{\sqrt{3}}}$ is fixed by the electron--hole pair density $n$. The self-consistent calculations are carried out on a $12\times12$ mBZ mesh at zero temperature, and the total energies are checked against finite-size scaling of the mesh.

\section{Ewald summation and total energy evaluation}
\label{app:ewald}

The total energy density is decomposed as $\mathcal{E}_{\mathrm T}=\mathcal{E}_{\mathrm H}+\mathcal{E}_{\mathrm{kin}}^{c}+\mathcal{E}_{\mathrm F}^{c}$, where $\mathcal{E}_{\mathrm H}$ is the classical electrostatic (Hartree) energy density of the complete charge-neutral electron--hole bilayer. Because each layer separately carries a net charge, its long-range Coulomb energy is not finite in isolation; only the combined neutral bilayer has a well-defined electrostatic energy, which we therefore evaluate as a single Ewald summation. The BLG carrier density and the localized $f$-hole density are written in reciprocal space as
\begin{equation}
\rho_c(\mathbf{r}) = \sum_{\bG}\rho_c(\bG)\,e^{i\bG\cdot\mathbf{r}},\qquad
\rho_f(\mathbf{r}) = \sum_{i,\bm{\tau}}\delta(\mathbf{r}-\mathbf{R}_i - \bm{\tau}) = \sum_{\bG}\rho_f(\bG)\,e^{i\bG\cdot\mathbf{r}},
\end{equation}
with $\rho_f(\bG)=\Omega_{\mathcal C}^{-1}\sum_{\bm{\tau}}e^{-i\bG\cdot\bm{\tau}}=S_{\mathcal C}(\bG)/\Omega_{\mathcal C}$, where $\Omega_{\mathcal C}$ is the supercell area and $S_{\mathcal C}$ the structure factor defined in the main text. The intralayer Hartree energy of the point-like $f$-holes cannot be summed directly in reciprocal space because the bare potential decays only as $V(\bG)\propto G^{-1}$. We therefore split the Coulomb kernel into rapidly convergent short- and long-range parts using a smooth screening function with tuning parameter $\alpha$,
\begin{equation}
V(\mathbf{r}) = V_s(\mathbf{r}) + V_l(\mathbf{r}),\quad
V_s(\mathbf{r}) = \frac{e^2\mathrm{erfc}(\alpha r)}{4\pi\varepsilon_0\varepsilon\, r},\quad
V_l(\mathbf{r}) = \frac{e^2\mathrm{erf}(\alpha r)}{4\pi\varepsilon_0\varepsilon\, r},\quad
V_l(\bG) = \frac{e^2\,\mathrm{erfc}\!\left(\tfrac{G}{2\alpha}\right)}{2\varepsilon_0\varepsilon\, G}.
\end{equation}
The $f$-hole Hartree energy density then separates into $\mathcal{E}_{\mathrm{H}}^{ff}=\mathcal{E}^{ff}_{\mathrm{H},s}+\mathcal{E}^{ff}_{\mathrm{H},l}$, with a real-space part summed within a cutoff $R_{\mathrm{cut}}$ and a reciprocal-space part,
\begin{align}
\mathcal{E}^{ff}_{\mathrm{H},s} &= \frac{1}{2\Omega_{\mathcal C}}\sum_{\bm{\tau}}\sum_{i,\bm{\tau}'}{}'\,
V_s(\mathbf{r}_i+\bm{\tau}'-\bm{\tau}) - \frac{1}{2\Omega_{\mathcal C}^2}V_s(\bG=\bm{0})\,N_{\bm{\tau}}^2,
\label{eq:eff-s}\\
\mathcal{E}^{ff}_{\mathrm{H},l} &= \frac{1}{2}\sum_{\bG\neq\bm{0}}|\rho_f(\bG)|^2\, V_l(\bG)-\frac{1}{2\Omega_{\mathcal C}}V_l(\mathbf{r}=\bm{0})\,N_{\bm{\tau}},
\label{eq:eff-l}
\end{align}
where $N_{\bm{\tau}}$ is the number of localized $f$-holes per supercell and the prime excludes the self-term. The subtractions in Eqs.~\eqref{eq:eff-s} and \eqref{eq:eff-l} remove, respectively, the divergent $\bG=\bm 0$ macroscopic energy of a uniformly charged layer and the point charge self-interaction, rendering the result independent of $\alpha$. The total Hartree energy density of the bilayer is
\begin{align}
\mathcal{E}_{\mathrm H} &= \mathcal{E}^{ff}_{\rm H}+\mathcal{E}^{cc}_{\rm H} + \mathcal{E}^{fc}_{\rm H},\\
\mathcal{E}^{cc}_{\rm H} &= \frac{e^2}{2}\sum_{\bG\neq\bm{0}}
\frac{1}{2\varepsilon_0\varepsilon\, G}\,|\rho_c(\bG)|^2,\\
\mathcal{E}^{fc}_{\rm H} &= -\frac{e^2}{\Omega_{\mathcal C}}\sum_{\bG\neq\bm{0}}\sum_{\bm{\tau}}
\frac{e^{-G d_0}}{2\varepsilon_0\varepsilon\, G}\,\rho_c(\bG)\,e^{i\bG\cdot\bm{\tau}},
\end{align}
where the interlayer term carries the form factor $e^{-G d_0}$ set by the BLG--MI separation $d_0$. Combining $\mathcal{E}_{\mathrm H}$ with the kinetic and Fock energy density of the converged Hartree Fock solution gives the total energy density which we compare across the triangular, honeycomb, and kagome charge orders at fixed density $n$ to determine the ground-state phase diagram.
The corresponding total energy per electron--hole pair is $\bar E_{\rm T}^{\mathcal C}=E_{\rm T}^{\mathcal C}/N_b$, where $E_{\rm T}^{\mathcal C}=\Omega_{\mathcal C}\mathcal E_{\rm T}^{\mathcal C}$ is the total energy per supercell.

\twocolumngrid

\bibliography{references}


\clearpage

\makeatletter
\counterwithout{equation}{section}
\def\theequation@prefix{}
\makeatother
\setcounter{equation}{0}
\setcounter{figure}{0}
\setcounter{table}{0}
\renewcommand{\theequation}{S\arabic{equation}}
\renewcommand{\thefigure}{S\arabic{figure}}
\renewcommand{\thetable}{S\arabic{table}}

\newcounter{smsec}
\renewcommand{\thesmsec}{S\arabic{smsec}}
\newcommand{\smsection}[1]{%
  \refstepcounter{smsec}%
  \section*{Note \thesmsec: #1}%
}

\onecolumngrid
\begin{center}
{\large\bfseries Supplemental Material for}\\[4pt]
{\large\bfseries ``Emergence of Topological Electron Crystals in Bilayer Graphene--Mott Insulator Heterostructures''}\\[8pt]
Wangqian Miao, Tianyu Qiao, Xue-Yang Song, Yinghai Xu, Yiwei Chen, Lei Wang, and Xi Dai
\end{center}
\vspace{8pt}

\smsection{Other charge transfer direction}
\label{sec:S-transfer}

\begin{figure}[!hb]
\centering
\includegraphics[width=0.8\textwidth]{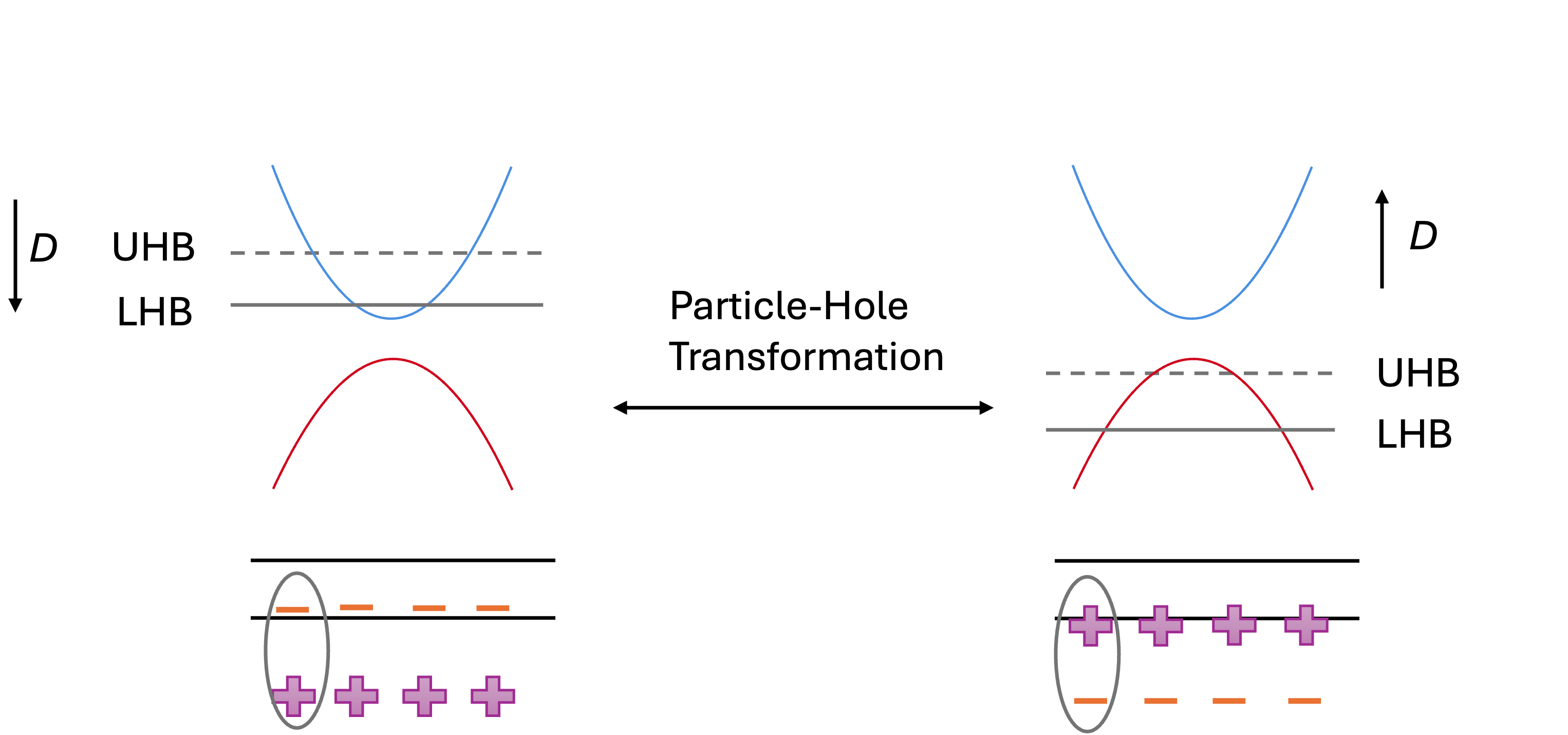}
\caption{\label{fig:transfer}
Schematic representation of charge transfer mechanisms. The electronic states of the Mott insulator (MI) are depicted as flat bands---the Upper Hubbard Band (UHB) and Lower Hubbard Band (LHB)---while the bilayer graphene (BLG) is represented by dispersive parabolic bands. Blue and red shading denote electron and hole sectors, respectively. (a) Case 2: Under a negative displacement field, electrons transfer from the MI-LHB to the BLG dispersive bands. (b) Case 3: Under a reversed displacement field, electrons transfer from the BLG to the MI-UHB. The two cases are related via an approximate particle-hole symmetry, illustrating the tunable nature of the interface doping through external fields.}
\end{figure}

To elucidate the doping mechanisms across the heterostructure interface, we analyze the charge transfer processes for Case 2 and Case 3, as defined in Table~\ref{tab:chern} of the main text. The electronic structure, illustrated schematically in Fig.~\ref{fig:transfer}, consists of the flat Upper and Lower Hubbard Bands (UHB/LHB) of the Mott insulator (MI) layer and the tunable dispersive bands of the bilayer graphene (BLG).

In Case 2, the electrons spontaneously transfer from the occupied LHB into the BLG, resulting in a hole-doped Mott layer and an electron-doped graphene layer. Conversely, Case 3 represents the conjugate scenario where the displacement field is reversed. Here, electrons transfer from the BLG into the empty UHB. This results in electron doping of the Mott layer and hole-doped BLG. Crucially, Case 2 and Case 3 are connected by an approximate particle-hole transformation. The same analysis can be performed for Cases 1 and 4. For simplicity, we mainly study Cases 1 and 2 in this paper.

\smsection{Single-particle analysis of the band topology}
\label{sec:S-topo}

\begin{figure}[t]
\centering
\includegraphics[width=0.95\textwidth]{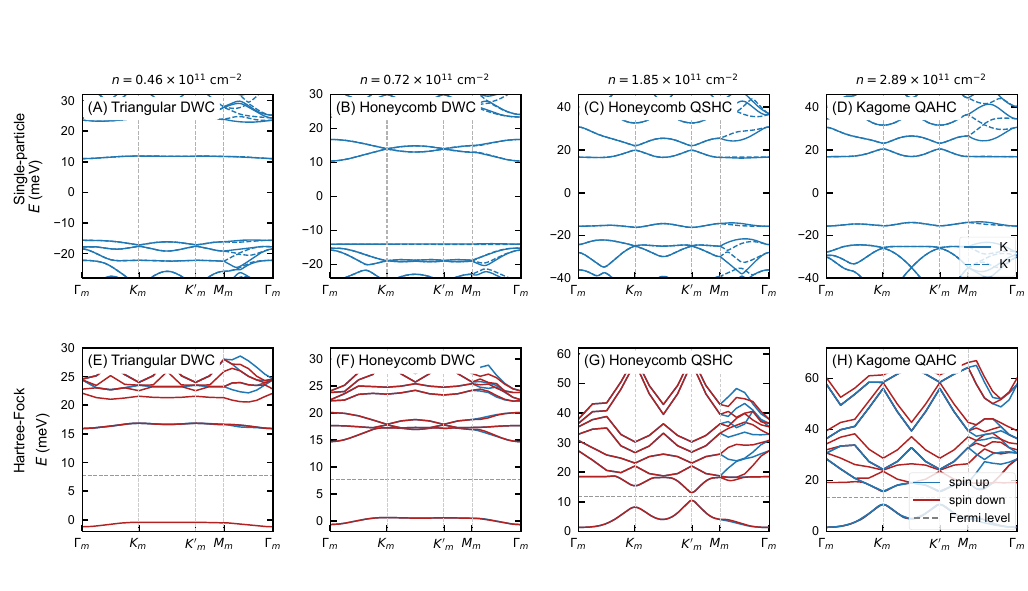}
\caption{\label{fig:bandsSM}
Band structures of the four states shown in Fig.~\ref{fig:density} of the main text. The top row shows the noninteracting single-particle bands; the bottom row shows the self-consistent Hartree Fock bands for the same parameters and fillings. From left to right, the panels correspond to the triangular DWC, honeycomb DWC, honeycomb QSHC, and kagome QAHC states at average densities $n=0.46$, $0.72$, $1.85$, and $2.89\times10^{11}\,\mathrm{cm}^{-2}$, respectively. In the top row, solid and dashed curves denote the two valley sectors $K$ and $K'$. In the bottom row, blue and red curves denote the spin-up and spin-down Hartree Fock bands, respectively, and the horizontal dashed line marks the chemical potential. The interaction-driven reconstruction of the low-energy bands is most pronounced in the honeycomb QSHC and kagome QAHC regimes.}
\end{figure}

In this section, we explain the single particle band topology of BLG under the superlattice Coulomb potential using perturbation theory. This analysis holds when the Coulomb potential does not invert the gap opened by the displacement field, so that it can be treated as a perturbation on the original BLG Hamiltonian. In the main text, the electron--hole attraction makes $U_{\mathcal C}(\mathbf G)$ negative for a triangular $f$-hole lattice. For the perturbative topology analysis it is useful to separate this common attractive sign from the lattice-geometry phase by defining
\begin{equation}
V_{\mathcal C}(\mathbf G)\equiv -U_{\mathcal C}(\mathbf G),\quad
U_{\mathcal C}(\mathbf r)=-\sum_{\bG} V_{\mathcal C}(\bG)\, e^{i\bG\cdot \mathbf r}.
\end{equation}
Projecting this potential onto a single Bloch band with Bloch functions $|\chi(\bk)\rangle$, the effective Hamiltonian reads
\begin{equation}
H(\bk)=\sum_{\bG}\varepsilon_{\bk+\bG}\, c^{\dagger}_{\bk+\bG}c_{\bk+\bG}
-\sum_{\bG}V_{\mathcal C}(\bG)\,\Lambda_{\bk+\bG;\bk}\, c^{\dagger}_{\bk+\bG}c_{\bk},
\end{equation}
where $\varepsilon_\bk$ is the original band dispersion and $\Lambda_{\bk';\bk}=\langle \chi(\bk')|\chi(\bk)\rangle$ is the form factor encoding the quantum geometry of the original band.

To analyze the topology of the minibands, we work in the original Bloch basis at the three momenta connected by $C_3$ rotation,
\begin{equation}
\mathcal B(\bq_{0})=\{\,|u_0\rangle\equiv|\chi(\bq_{0})\rangle,\;
|u_1\rangle\equiv|\chi(R\bq_{0})\rangle,\;
|u_{2}\rangle\equiv|\chi(R^{2}\bq_{0})\rangle\,\},
\end{equation}
with $R$ the $2\pi/3$ rotation and $\bq_0=\kappa,\kappa'$ the high-symmetry points in the mini Brillouin zone (mBZ). In this basis the effective Hamiltonian takes the circulant form
\begin{equation}
H=\begin{pmatrix}
a_{0} & a_{2} & a_{1}\\
a_{1} & a_{0} & a_{2}\\
a_{2} & a_{1} & a_{0}
\end{pmatrix},
\end{equation}
with $a_{0}=\varepsilon_{\bq_{0}}$, $a_{1}=-V_{\mathcal C}(G_{1})\langle u_{1}|u_{0}\rangle$ and $a_{2}=a_{1}^{*}$. The normalized eigenvectors are
\begin{equation}
|\phi_{\nu}\rangle=\frac{1}{\sqrt{3}}\Big(|u_{0}\rangle+\omega^{-\nu}|u_{1}\rangle+\omega^{-2\nu}|u_{2}\rangle\Big),
\end{equation}
with $\nu=0,1,2$, $\omega=e^{2\pi i/3}$, and eigenvalues
\begin{equation}
E_{\nu}(\bq_{0})=\varepsilon_{\bq_{0}}
+2|a_{1}|\cos\!\Big(\varphi(\bq_{0})+\tfrac{2\pi\nu}{3}\Big),
\end{equation}
where $\varphi(\bq_{0})=\arg(a_{1})$. The $C_{3}$ operator acts on the basis as
\begin{equation}
U_{3}|u_{0}\rangle=S_{0}|u_{1}\rangle,\quad
U_{3}|u_{1}\rangle=S_{1}|u_{2}\rangle,\quad
U_{3}|u_{2}\rangle=S_{2}|u_{0}\rangle,
\end{equation}
where $S_{j}=\langle u_{j+1}|U_{3}|u_{j}\rangle$ are sewing phases satisfying $S_{0}S_{1}S_{2}=1$. Gauge freedom allows us to homogenize $S_{j}$, such that
\begin{equation}
U_{3}|\phi_{\nu}\rangle=\omega^{\nu}B\,|\phi_{\nu}\rangle,\qquad |B|=1.
\end{equation}
Choosing $\Gamma$ as a reference point, we define
\begin{equation}
M_{\Gamma}=\langle\chi(\Gamma)|U_{3}|\chi(\Gamma)\rangle,\qquad
M_{\kappa}=\langle\chi(\kappa)|U_{3}|\chi(\kappa)\rangle,\qquad
M_{\kappa'}=\langle\chi(\kappa')|U_{3}|\chi(\kappa')\rangle.
\end{equation}
The $C_{3}$ eigenvalues for the occupied band then read
\begin{equation}
\lambda(\Gamma)=M_{\Gamma},\qquad
\lambda(\kappa)=\omega^{\nu_{\kappa}}\frac{M_{\kappa}}{M_{\Gamma}},\qquad
\lambda(\kappa')=\omega^{\nu_{\kappa'}}\frac{M_{\kappa'}}{M_{\Gamma}}.
\end{equation}
The product rule $M_{\kappa}M_{\kappa'}=M_{\Gamma}$ ensures cancellation of material phases, leading to
\begin{equation}
e^{2\pi i C_{3}/3}=\lambda(\Gamma)\lambda(\kappa)\lambda(\kappa')
=\omega^{\nu_{\kappa}+\nu_{\kappa'}},\qquad
C_{3}\equiv \nu_{\kappa}+\nu_{\kappa'}\;\;(\mathrm{mod}\;3).
\end{equation}
Hence the Chern number is completely determined by the indices $\nu_{\kappa},\nu_{\kappa'}$ of the lowest band. The choice of the lowest eigenvalue is fixed by the loop phase [see Ref.~\cite{crepel_efficient_2025}],
\begin{align}
\nu(\bq_0)&\equiv
\left\lfloor \tfrac{\pi-\Phi^{\circlearrowleft}_{\bq_0}}{2\pi}\right\rfloor,\\
\Phi^{\circlearrowleft}_{\bq_{0}}
&=\arg\!\left[
V_{\mathcal C}(G_{1})V_{\mathcal C}(G_{2})V_{\mathcal C}(G_{3})
\;M_{\bq_{0}}M_{R\bq_{0}}M_{R^{2}\bq_{0}}
\right],
\end{align}
where $G_{1,2,3}$ are the three reciprocal vectors connecting the loop
$\bq_{0}\!\to\!R\bq_{0}\!\to\!R^{2}\bq_{0}\!\to\!\bq_{0}$, and
$M_{\bq}=\langle \chi(\bq)|U_{3}|\chi(\bq)\rangle$. Under a local gauge transformation $|\chi(\bq)\rangle\to e^{i\alpha(\bq)}|\chi(\bq)\rangle$, each $M_{\bq}$ acquires a phase $e^{i(\alpha(\bq)-\alpha(R\bq))}$, which cancels in the product
$M_{\bq_{0}}M_{R\bq_{0}}M_{R^{2}\bq_{0}}$. Hence $\Phi^{\circlearrowleft}_{\bq_{0}}$ is manifestly gauge invariant, and the Chern number modulo $3$ is
\begin{equation}
C_{3}\equiv \nu(\kappa)+\nu(\kappa')\;\;(\mathrm{mod}\;3).
\end{equation}
For a very long superlattice period, the loop phase reduces to
\begin{equation}
\Phi^{\circlearrowleft}_{\kappa}\simeq 3\theta+\tfrac{1}{2}\phi_{B},\qquad
\Phi^{\circlearrowleft}_{\kappa'}\simeq -3\theta+\tfrac{1}{2}\phi_{B},
\end{equation}
with $\theta=\arg[V_{\mathcal C}(\bG)]$ and $\phi_{B}$ the total Berry flux in the mBZ. The Chern number is then
\begin{equation}
C_{3}\equiv
\left\lfloor \tfrac{\pi-(3\theta+\phi_{B}/2)}{2\pi}\right\rfloor
+ \left\lfloor \tfrac{\pi-(-3\theta+\phi_{B}/2)}{2\pi}\right\rfloor
\;\;(\mathrm{mod}\;3),
\end{equation}
giving $C_{3}=0$ for triangular potentials ($\theta=0$, $|\phi_{B}|<2\pi$), and $C_{3}=\pm 1$ for honeycomb potentials ($\theta=\pi$), with the sign set by $\mathrm{sgn}(\phi_{B})$.

The above single-particle analysis is controlled under the following conditions:
\begin{enumerate}
\item Weak superlattice Coulomb potential, such that only the $C_{3}$-invariant corners are resonantly coupled.
\item Band isolation: the target miniband remains separated by gaps from neighboring bands across the mBZ, so that no gap closes upon increasing $|U_{\mathcal C}(\bG)|$.
\end{enumerate}
This analysis explains why the first conduction band of the BLG--MI bilayer is topologically trivial for the triangular lattice but nontrivial for the honeycomb and kagome lattices. Furthermore, the sign of $\phi_B$ is set by the original Berry-curvature distribution of BLG under the displacement field. The downfolded two-band model for BLG under a displacement field is
\begin{equation}
H_{\xi}(\bk)=
\begin{pmatrix}
\Delta & -\dfrac{(\xi k_x+i k_y)^2}{2m} \\[6pt]
-\dfrac{(\xi k_x-i k_y)^2}{2m} & -\Delta
\end{pmatrix},
\end{equation}
with $m=\gamma_1/(2v_F^2)$, and the corresponding Berry curvature of each band is
\begin{equation}
\Omega_{\pm}^\xi(k)=
\pm\xi\,
\frac{\Delta\,k^2/2m^2}
{\left[\Delta^2+\left(\dfrac{k^2}{2m}\right)^2\right]^{3/2}}.
\end{equation}
The displacement-field direction thus controls the sign of the Berry curvature, which determines the sign of $\phi_B$. Combining these observations, the Table~\ref{tab:chern} results of the main text can be fully understood within perturbation theory and are confirmed by the numerics in our parameter regime. The reconstructed single-particle bands and the corresponding self-consistent Hartree Fock quasiparticle bands for the four representative states of Fig.~\ref{fig:density} of the main text are shown in Fig.~\ref{fig:bandsSM}.

\smsection{Classical image charge limit}
\label{sec:S-classical}

\begin{figure}[t]
\centering
\includegraphics[width=0.9\textwidth]{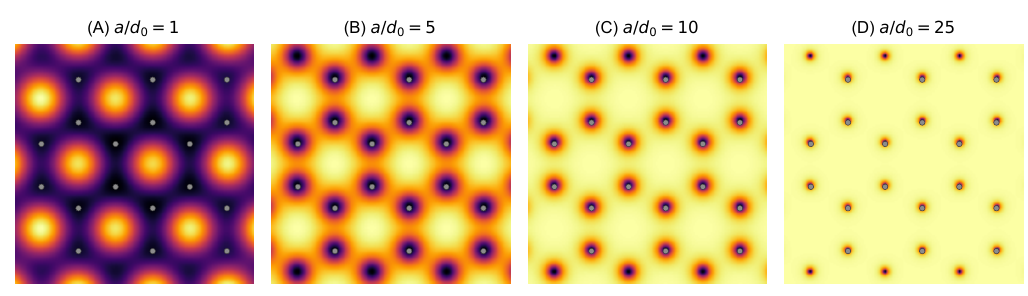}
\caption{\label{fig:imageSM}
Induced charge distribution on the grounded metallic plate when isolated charges of opposite sign are arranged in a honeycomb lattice. The ratio between the lattice constant $a$ and the interlayer distance $d_0$ is (A) $1$, (B) $5$, (C) $10$ and (D) $25$. Gray dots label the isolated charges on the other layer.}
\end{figure}

To illustrate the classical limit of the electron--hole bilayer problem with interlayer distance $d_0$, we consider a periodic array of charges located at positions $\mathbf{R}_i+\bm{\tau}$ in the $x$--$y$ plane, at a vertical distance $d_0$ below a grounded metallic plate (located at $z=0$). The electrostatic potential at a point $(\mathbf{r},z)$ above the plate is obtained using the method of images,
\begin{equation}
\Phi(\mathbf{r},z) =
\sum_{i,\bm{\tau}} \left(
\frac{q}{4\pi\varepsilon_0 \sqrt{|\mathbf{r}-\mathbf{R}_i-\bm{\tau}|^2+(z-d_0)^2}}
-
\frac{q}{4\pi\varepsilon_0 \sqrt{|\mathbf{r}-\mathbf{R}_i-\bm{\tau}|^2+(z+d_0)^2}}
\right).
\end{equation}
The first term is the direct Coulomb potential of the array of charges, while the second corresponds to the image charges introduced to satisfy the metallic boundary condition at $z=0$. The perpendicular electric field just above the plate is
\begin{equation}
E_z(\mathbf{r}, z=0^+)
= -\sum_{i,\bm{\tau}}
\frac{2 q d_0}{4\pi\varepsilon_0 \left(|\mathbf{r}-\mathbf{R}_i-\bm{\tau}|^2+d_0^2\right)^{3/2}},
\end{equation}
where the direct charges and their images contribute equally. Finally, the induced charge density on the plate follows from Gauss's law,
\begin{equation}
\rho(\mathbf{r}) = \varepsilon_0 E_z(\mathbf{r},z=0^+)=-
\sum_{i,\bm{\tau}}
\frac{q d_0}{2\pi \left(|\mathbf{r}-\mathbf{R}_i-\bm{\tau}|^2+d_0^2\right)^{3/2}}.
\end{equation}
This describes the spatial modulation of induced surface charge on the metallic plate due to the periodic arrangement of charges in the lower layer. It represents the classical electrostatic limit of the electron--hole bilayer system, in which the interlayer coupling reduces to image-charge interactions when the lattice constant $a\gg d_0$. We numerically solve the induced charge distribution on the plate for charges of opposite sign forming a honeycomb lattice beneath, as shown in Fig.~\ref{fig:imageSM}.

\smsection{Additional phase diagrams}
\label{sec:S-phase}

\begin{figure}[t]
\centering
\includegraphics[width=0.42\textwidth]{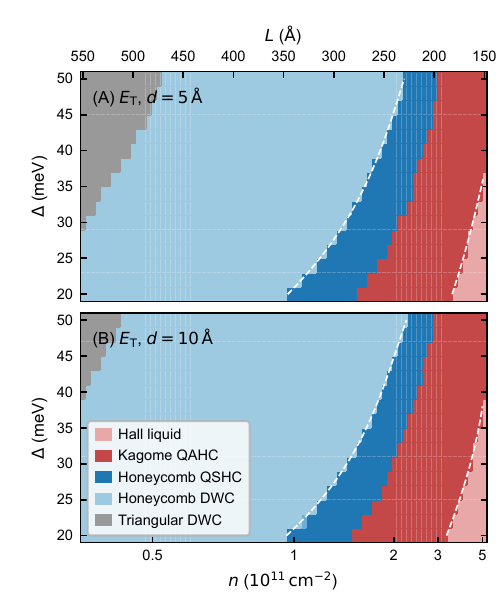}
\caption{\label{fig:phaseSM}
Phase diagrams computed as in Fig.~\ref{fig:phase} of the main text for different interlayer distances $d_0$, with dielectric constant $\varepsilon=8$.}
\end{figure}

\begin{figure}[t]
\centering
\includegraphics[width=0.8\textwidth]{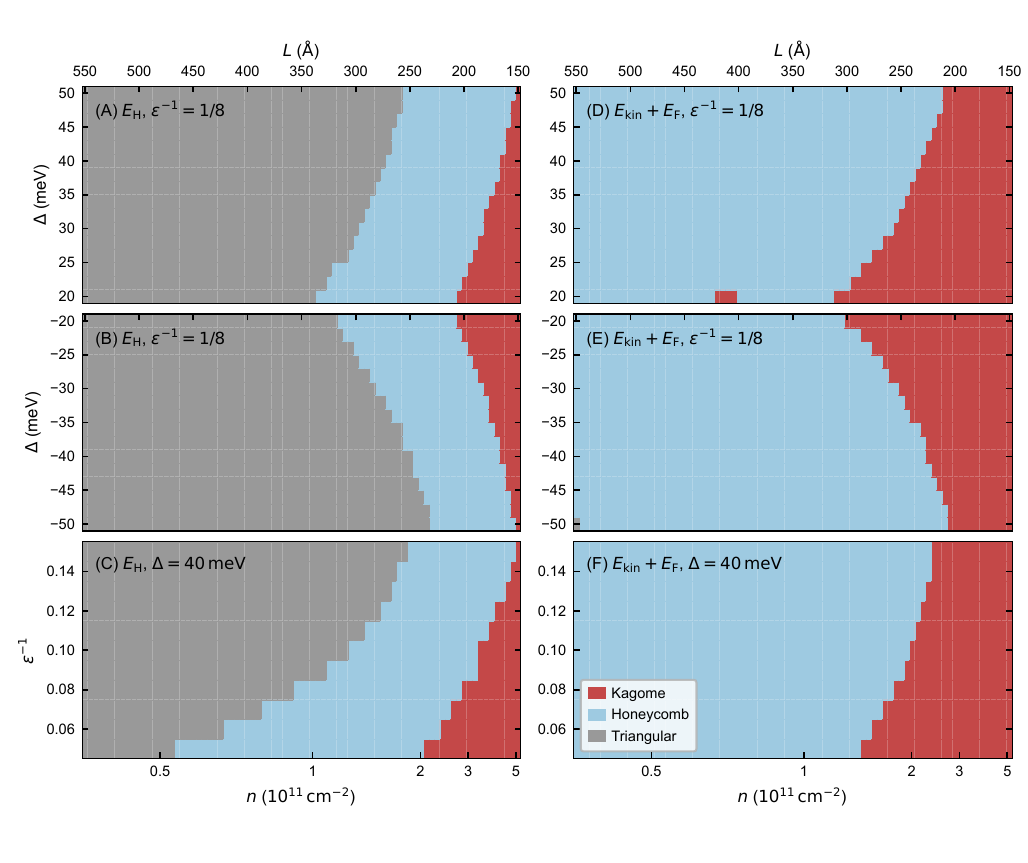}
\caption{\label{fig:phaseHartree}
Phase diagrams computed as in Fig.~\ref{fig:phase} of the main text with (A-C) only Hartree energy, (D-F) only kinetic and Fock energy. $d_0$ is fixed as $1$ \AA.}
\end{figure}

In this section we present additional phase diagrams obtained by varying the interlayer distance $d_0$; see Fig.~\ref{fig:phaseSM}. These Hartree Fock calculations are performed on a $12\times12$ $k$-mesh at zero temperature. The phase diagrams share the same qualitative features as those in the main text.
In Fig.~\ref{fig:phaseHartree}, we also present the phase diagram obtained by retaining only the Hartree energy, using the same parameters as in Figs.~\ref{fig:phase}(A), (C), and (D) of the main text. We see that the triangular configuration expands and dominates in the low-density regime, whereas the honeycomb configuration yields a lower Hartree energy at higher densities. When considering the quantum energies (kinetic energy plus Fock energy), the honeycomb configuration dominates.

\end{document}